\documentclass[acmlarge]{acmart}

\AtBeginDocument{%
  \providecommand\BibTeX{{%
    \normalfont B\kern-0.5em{\scshape i\kern-0.25em b}\kern-0.8em\TeX}}}

\copyrightyear{2020} 
\acmYear{2020} 
\setcopyright{rightsretained} 
\acmConference[JCDL '20]{Proceedings of the ACM/IEEE Joint Conference on Digital Libraries in 2020}{August 1--5, 2020}{Virtual Event, China}
\acmBooktitle{Proceedings of the ACM/IEEE Joint Conference on Digital Libraries in 2020 (JCDL '20), August 1--5, 2020, Virtual Event, China}\acmDOI{10.1145/3383583.3398561}
\acmISBN{978-1-4503-7585-6/20/06}

\setcopyright{none}

\usepackage{cleveref}
\usepackage[absolute,overlay]{textpos}

\begin{document}

\title{Newsalyze: Enabling News Consumers to Understand Media Bias}

\author{Felix Hamborg${}^{1}$,
	Anastasia Zhukova${}^{2}$,
	Karsten Donnay${}^{3,4}$, 
	Bela Gipp${}^{2}$
}
\newcommand{\authorsclean}{Felix Hamborg, Anastasia Zhukova, Karsten Donnay, Bela Gipp}
\affiliation{
	\institution{${}^{1}$Dept. of Computer and Information Science, University of Konstanz, Germany, felix.hamborg@uni-konstanz.de}
}
\affiliation{
	\institution{${}^{2}$Data \& Knowledge Engineering Group, University of Wuppertal, Wuppertal, Germany, \{last\}@uni-wuppertal.de}
}

\affiliation{
	\institution{${}^{3}$Dept. of Political Science, University of Zurich, Switzerland, karsten.donnay@uzh.ch}
}

\affiliation{
	\institution{${}^{4}$Dept. of Politics and Public Administration, University of Konstanz, Germany, karsten.donnay@uni-konstanz.de}
}

\fancyhead{}
\renewcommand{\shortauthors}{Hamborg, Zhukova, Donnay, and Gipp}

\begin{abstract}
  News is a central source of information for individuals to inform themselves on current topics. Knowing a news article's slant and authenticity is of crucial importance in times of ``fake news,'' news bots, and centralization of media ownership. We introduce \emph{Newsalyze}, a bias-aware news reader focusing on a subtle, yet powerful form of media bias, named bias by word choice and labeling (WCL). WCL bias can alter the assessment of entities reported in the news, e.g., ``freedom fighters'' vs. ``terrorists.'' At the core of the analysis is a neural model that uses a news-adapted BERT language model to determine target-dependent sentiment, a high-level effect of WCL bias. While the analysis currently focuses on only this form of bias, the visualizations already reveal patterns of bias when contrasting articles (overview) and in-text instances of bias (article view).
\end{abstract}


\maketitle

\begin{textblock*}{\paperwidth}(0cm,1cm) 
	\centering
	\noindent
	This is a preprint. The accepted manuscript can be found at: \\ \url{https://doi.org/10.1145/3383583.3398561}
\end{textblock*}

\section{Introduction and Related Work}
People rely on the news to inform themselves on current topics and events. Especially news articles, which the public commonly deems most trustworthy \cite{Urban1999ExaminingPress}, are a central part of individual and societal opinion formation and decision making. Media bias, e.g., slanted or biased news coverage, thus can have severe effects on democratic processes \cite{meyrowitz1986no}. A subtle, yet powerful form of media bias is bias by word choice and labeling (WCL), which occurs when news authors sway readers' perception of persons, actions, or other semantic concepts by using different terms or phrases to refer to the concepts, e.g., "undocumented immigrant" vs. "illegal alien." Previous works have struggled to automatically identify WCL bias \cite{balahur2013sentiment, godbole2007large, Hamborg2019a}, mainly due its implicitness, subjectivity, and high context dependence \cite{Card2015TheIssues}, requiring actual understanding of the text at hand. However, the advent of deep learning and language models, such as BERT, has led to a significant leap towards natural language understanding (NLU), thereby strongly improving the performance in many tasks deemed traditionally as difficult \cite{Wang2018GLUE:Understanding}. 

To our knowledge, there is no news reader that enables bias comparison of articles reporting on the same topic and exploration of bias instances within an article. More importantly, no bias-related approaches leverage most recent advancements in NLU, which could help to significantly improve the detection performance of biases that could not be addressed well before. We propose \emph{Newsalyze}, a news reader that analyzes and visualizes WCL bias in news articles. The prototype currently focuses on visualizing a high-level effect of WCL bias and determines whether a target, i.e., a semantic concept, is portrayed positively or negatively within a sentence.

\section{System and User's Workflow}
The system performs a five task workflow (cf. \cite{Hamborg2019a,Hamborg2019}): article gathering, preprocessing, target concept analysis, frame identification, and visualization. For \textit{article gathering}, we crawl and extract news articles, currently for given a set of user-defined URLs \cite{Hamborg2017a} for each topic. We then perform state-of-the-art NLP \textit{preprocessing} using Stanford CoreNLP. \textit{Target concept analysis} finds and resolves semantic concepts, such as persons or countries, across each topic's articles, going beyond regular coreference resolution by finding also broadly or abstractly defined as well as contrarily mentioned concepts, such as "freedom fighters" vs. "terrorists" \cite{Hamborg2019a}. \textit{Frame identification} determines how concepts are portrayed in their mentions, e.g., ranging from sentiment polarity (positive or negative) to fine-grained framing effects, e.g., whether a person is portrayed as being "competent", "weak" or "aggressive" \cite{Hamborg2019a}. Identifying frames is a challenging task, for human coders \cite{Card2015TheIssues} as well as for previous automated approaches, which either yield mixed results if aiming to find universally valid frames \cite{Hamborg2019a} or are specialized to only one or a few topics \cite{Greussing2017ShiftingCrisis}. Thus, we currently focus on targeted sentiment, which is a high-level effect of WCL bias but also a universal perception dimension. To achieve state of-the-art performance in target-dependent sentiment classification (TSC) on news articles, we use NewsTSC, a BERT-based neural model \cite{Hamborg2020MediaLabeling}.

\begin{figure}
	\includegraphics[width=\textwidth]{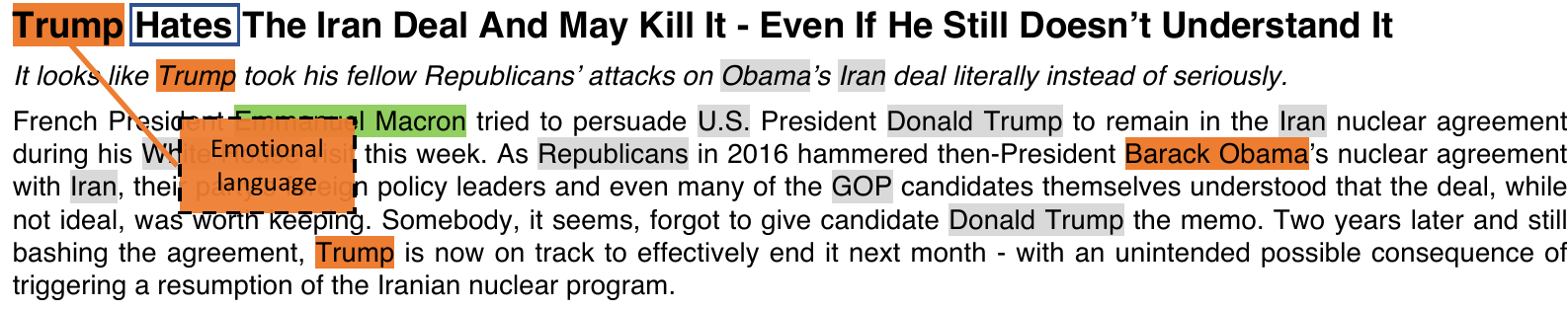}
	\caption{Newsalyze's article view highlights mentions of semantic concepts, such as persons, according to their target-dependent sentiment, a high-level effect of bias by word choice and labeling (green: positive, red: negative).}
	\label{fig:teaser}
\end{figure}

\begin{figure}[h]
	\centering
	\includegraphics[width=.5\linewidth]{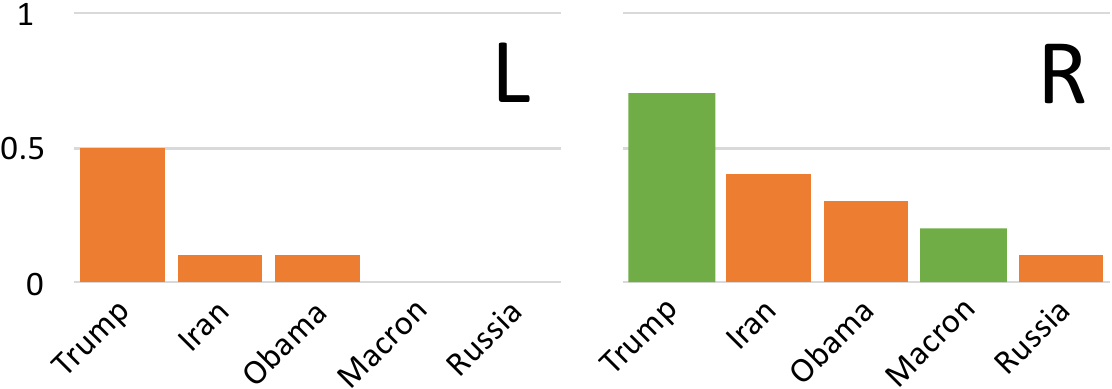}
	\caption{Framing histograms of a topic's most frequent semantic concepts, shown for a left-slanted (L) and a right-slanted (R) article. Each bar's height represents the frequency of its concept, the color aggregated positive (green) or negative (red) sentiment of the concept. \label{fig:framinghist}}
	\Description{Framing histogram of a topic's most frequent semantic concepts.}
\end{figure}

Lastly, the system visualizes the identified instances of WCL bias using two visualizations, which follow the overview first, details on demand mantra \cite{Shneiderman1996}. First, an \textit{overview}, similar to the overview offered by news aggregators such as Google News, shows current topics and for each topic a selection of articles reporting on it. Newsalyze's overview enables users to efficiently compare how articles portray the topic's most important concepts: besides each article snippet, the visualization shows a histogram representing the article's normalized sentiment of the topic's most frequent concepts. \Cref{fig:framinghist} shows histograms of two articles reporting on the Iran deal topic published by HuffPost (left-slanted outlet) and Breitbart (right) in April 2018. Second, an \textit{article-view} helps users to understand WCL bias while reading an article, e.g., by visually highlighting concept mentions as to the bias categories identified for them on sentence-level. \Cref{fig:teaser} shows an excerpt of the left-slanted article.

Using the overview, users can quickly understand current topics. In contrast to common news aggregators, the overview is bias-aware: its framing histogram shown besides each article snippet enables users to quickly compare how important actors are portrayed across the topic. For example, \Cref{fig:framinghist} shows aggregated polarities of Trump and other most frequent NEs of the Iran deal topic. The visual comparison immediately reveals that Trump is portrayed rather negatively in the left outlet but strongly positively in the right outlet. In common news aggregators, users would have to read whole articles to come to this conclusion. Lastly, the article-view aids user to understand bias simply while reading the article, because, for example, phrases of WCL bias are visually highlighted.

\section{Conclusion and Future Work}
Newsalyze is the first bias-aware news reader that supports the full news consumption process, from getting an overview of current topics as well as reading articles. By contrasting how a topic's actors are portrayed by each article, users can efficiently get an overview not only of the topic but also of the slant of each article. Afterward, when reading an article of interest, users are aided to see bias with the help of in-text bias markers. The system currently analyzes and visualizes a high-level effect of bias by word choice and labeling (WCL), i.e., target-dependent sentiment. In the future, we plan to devise and train a neural model to additionally classify more fine-grained perception dimensions, e.g., framing effects such as whether a person is portrayed as competent or incompetent. We also plan to classify causes of the identified WCL instances, e.g., the use of emotional language (see \Cref{fig:teaser}). We hope that in the future systems such as Newsalyze will help people to become aware of bias conveniently during their daily news consumption. The recently increased interest in this topic, not only in research communities but also in society, emphasizes the issue's importance.

\begin{acks}
The work described in this paper is partially funded by the WIN program of the Heidelberg Academy of Sciences and Humanities, financed by the Ministry of Science, Research and the Arts of the State of Baden-Wurttemberg, Germany. 
\end{acks}

\bibliographystyle{ACM-Reference-Format}
\bibliography{references}


\begin{thebibliography}{12}

%%% the Web address.

\ifx \showCODEN    \undefined \def \showCODEN     #1{\unskip}     \fi
\ifx \showDOI      \undefined \def \showDOI       #1{#1}\fi
\ifx \showISBNx    \undefined \def \showISBNx     #1{\unskip}     \fi
\ifx \showISBNxiii \undefined \def \showISBNxiii  #1{\unskip}     \fi
\ifx \showISSN     \undefined \def \showISSN      #1{\unskip}     \fi
\ifx \showLCCN     \undefined \def \showLCCN      #1{\unskip}     \fi
\ifx \shownote     \undefined \def \shownote      #1{#1}          \fi
\ifx \showarticletitle \undefined \def \showarticletitle #1{#1}   \fi
\ifx \showURL      \undefined \def \showURL       {\relax}        \fi
% invisible to TeX
\providecommand\bibfield[2]{#2}
\providecommand\bibinfo[2]{#2}
\providecommand\natexlab[1]{#1}
\providecommand\showeprint[2][]{arXiv:#2}

\bibitem[\protect\citeauthoryear{Balahur, Steinberger, Kabadjov, Zavarella, Van
  Der~Goot, Halkia, Pouliquen, and Belyaeva}{Balahur et~al\mbox{.}}{2013}]%
        {balahur2013sentiment}
\bibfield{author}{\bibinfo{person}{Alexandra Balahur}, \bibinfo{person}{Ralf
  Steinberger}, \bibinfo{person}{Mijail Kabadjov}, \bibinfo{person}{Vanni
  Zavarella}, \bibinfo{person}{Erik Van Der~Goot}, \bibinfo{person}{Matina
  Halkia}, \bibinfo{person}{Bruno Pouliquen}, {and} \bibinfo{person}{Jenya
  Belyaeva}.} \bibinfo{year}{2013}\natexlab{}.
\newblock \showarticletitle{{Sentiment analysis in the news}}.
\newblock \bibinfo{journal}{\emph{arXiv preprint arXiv:1309.6202}}
  (\bibinfo{year}{2013}).
\newblock


\bibitem[\protect\citeauthoryear{Card, Boydstun, Gross, Resnik, and Smith}{Card
  et~al\mbox{.}}{2015}]%
        {Card2015TheIssues}
\bibfield{author}{\bibinfo{person}{Dallas Card}, \bibinfo{person}{Amber~E.
  Boydstun}, \bibinfo{person}{Justin~H. Gross}, \bibinfo{person}{Philip
  Resnik}, {and} \bibinfo{person}{Noah~A. Smith}.}
  \bibinfo{year}{2015}\natexlab{}.
\newblock \showarticletitle{{The Media Frames Corpus: Annotations of Frames
  Across Issues}}. In \bibinfo{booktitle}{\emph{Proceedings of the 53rd Annual
  Meeting of the Association for Computational Linguistics and the 7th
  International Joint Conference on Natural Language Processing (Volume 2:
  Short Papers)}}. \bibinfo{publisher}{Association for Computational
  Linguistics}, \bibinfo{address}{Stroudsburg, PA, USA},
  \bibinfo{pages}{438--444}.
\newblock
\urldef\tempurl%
\url{https://doi.org/10.3115/v1/P15-2072}
\showDOI{\tempurl}


\bibitem[\protect\citeauthoryear{Godbole, Srinivasaiah, and Skiena}{Godbole
  et~al\mbox{.}}{2007}]%
        {godbole2007large}
\bibfield{author}{\bibinfo{person}{Namrata Godbole}, \bibinfo{person}{Manja
  Srinivasaiah}, {and} \bibinfo{person}{Steven Skiena}.}
  \bibinfo{year}{2007}\natexlab{}.
\newblock \showarticletitle{{Large-Scale Sentiment Analysis for News and
  Blogs}}.
\newblock \bibinfo{journal}{\emph{ICWSM}} \bibinfo{volume}{7},
  \bibinfo{number}{21} (\bibinfo{year}{2007}), \bibinfo{pages}{219--222}.
\newblock


\bibitem[\protect\citeauthoryear{Greussing and Boomgaarden}{Greussing and
  Boomgaarden}{2017}]%
        {Greussing2017ShiftingCrisis}
\bibfield{author}{\bibinfo{person}{Esther Greussing} {and}
  \bibinfo{person}{Hajo~G. Boomgaarden}.} \bibinfo{year}{2017}\natexlab{}.
\newblock \showarticletitle{{Shifting the refugee narrative? An automated frame
  analysis of Europe’s 2015 refugee crisis}}.
\newblock \bibinfo{journal}{\emph{Journal of Ethnic and Migration Studies}}
  \bibinfo{volume}{43}, \bibinfo{number}{11} (\bibinfo{date}{8}
  \bibinfo{year}{2017}), \bibinfo{pages}{1749--1774}.
\newblock
\showISSN{1369-183X}
\urldef\tempurl%
\url{https://doi.org/10.1080/1369183X.2017.1282813}
\showDOI{\tempurl}


\bibitem[\protect\citeauthoryear{Hamborg}{Hamborg}{2020}]%
        {Hamborg2020MediaLabeling}
\bibfield{author}{\bibinfo{person}{Felix Hamborg}.}
  \bibinfo{year}{2020}\natexlab{}.
\newblock \showarticletitle{{Media Bias, the Social Sciences, and NLP:
  Automating Frame Analyses to Identify Bias by Word Choice and Labeling}}. In
  \bibinfo{booktitle}{\emph{Proceedings of the 58th Annual Meeting of the
  Association for Computational Linguistics (ACL): Student Research Workshop}}.
  \bibinfo{publisher}{Association for Computational Linguistics},
  \bibinfo{pages}{1--9}.
\newblock


\bibitem[\protect\citeauthoryear{Hamborg, Meuschke, Breitinger, and
  Gipp}{Hamborg et~al\mbox{.}}{2017}]%
        {Hamborg2017a}
\bibfield{author}{\bibinfo{person}{Felix Hamborg}, \bibinfo{person}{Norman
  Meuschke}, \bibinfo{person}{Corinna Breitinger}, {and} \bibinfo{person}{Bela
  Gipp}.} \bibinfo{year}{2017}\natexlab{}.
\newblock \showarticletitle{{news-please: A Generic News Crawler and
  Extractor}}. In \bibinfo{booktitle}{\emph{Proceedings of the 15th
  International Symposium of Information Science}}. \bibinfo{publisher}{Verlag
  Werner H{\"{u}}lsbusch}, \bibinfo{pages}{218--223}.
\newblock


\bibitem[\protect\citeauthoryear{Hamborg, Zhukova, and Gipp}{Hamborg
  et~al\mbox{.}}{2019a}]%
        {Hamborg2019a}
\bibfield{author}{\bibinfo{person}{Felix Hamborg}, \bibinfo{person}{Anastasia
  Zhukova}, {and} \bibinfo{person}{Bela Gipp}.}
  \bibinfo{year}{2019}\natexlab{a}.
\newblock \showarticletitle{{Automated Identification of Media Bias by Word
  Choice and Labeling in News Articles}}. In \bibinfo{booktitle}{\emph{2019
  ACM/IEEE Joint Conference on Digital Libraries (JCDL)}}.
  \bibinfo{publisher}{IEEE}, \bibinfo{address}{Urbana-Champaign, IL, USA},
  \bibinfo{pages}{196--205}.
\newblock
\showISBNx{978-1-7281-1547-4}
\urldef\tempurl%
\url{https://doi.org/10.1109/JCDL.2019.00036}
\showDOI{\tempurl}


\bibitem[\protect\citeauthoryear{Hamborg, Zhukova, and Gipp}{Hamborg
  et~al\mbox{.}}{2019b}]%
        {Hamborg2019}
\bibfield{author}{\bibinfo{person}{Felix Hamborg}, \bibinfo{person}{Anastasia
  Zhukova}, {and} \bibinfo{person}{Bela Gipp}.}
  \bibinfo{year}{2019}\natexlab{b}.
\newblock \showarticletitle{{Illegal Aliens or Undocumented Immigrants? Towards
  the Automated Identification of Bias by Word Choice and Labeling}}.
\newblock In \bibinfo{booktitle}{\emph{Proceedings of the iConference 2019}}.
  \bibinfo{publisher}{Springer, Cham}, \bibinfo{address}{Washington, DC, USA},
  \bibinfo{pages}{179--187}.
\newblock
\showISBNx{978-3-030-15741-8}
\urldef\tempurl%
\url{https://doi.org/10.1007/978-3-030-15742-5{\_}17}
\showDOI{\tempurl}


\bibitem[\protect\citeauthoryear{Meyrowitz}{Meyrowitz}{1986}]%
        {meyrowitz1986no}
\bibfield{author}{\bibinfo{person}{Joshua Meyrowitz}.}
  \bibinfo{year}{1986}\natexlab{}.
\newblock \bibinfo{booktitle}{\emph{{No sense of place: The impact of
  electronic media on social behavior}}}.
\newblock \bibinfo{publisher}{Oxford University Press}.
\newblock


\bibitem[\protect\citeauthoryear{Shneiderman}{Shneiderman}{1996}]%
        {Shneiderman1996}
\bibfield{author}{\bibinfo{person}{B. Shneiderman}.}
  \bibinfo{year}{1996}\natexlab{}.
\newblock \showarticletitle{{The eyes have it: a task by data type taxonomy for
  information visualizations}}.
\newblock \bibinfo{journal}{\emph{Proceedings 1996 IEEE Symposium on Visual
  Languages}} (\bibinfo{year}{1996}), \bibinfo{pages}{336--343}.
\newblock
\showISBNx{0-8186-7508-X}
\showISSN{1049-2615}
\urldef\tempurl%
\url{https://doi.org/10.1109/VL.1996.545307}
\showDOI{\tempurl}


\bibitem[\protect\citeauthoryear{Urban}{Urban}{1999}]%
        {Urban1999ExaminingPress}
\bibfield{author}{\bibinfo{person}{Christine~D Urban}.}
  \bibinfo{year}{1999}\natexlab{}.
\newblock \bibinfo{booktitle}{\emph{{Examining Our Credibility: Perspectives of
  the Public and the Press}}}.
\newblock \bibinfo{publisher}{Asne Foundation}. 1--108 pages.
\newblock


\bibitem[\protect\citeauthoryear{Wang, Singh, Michael, Hill, Levy, and
  Bowman}{Wang et~al\mbox{.}}{2018}]%
        {Wang2018GLUE:Understanding}
\bibfield{author}{\bibinfo{person}{Alex Wang}, \bibinfo{person}{Amanpreet
  Singh}, \bibinfo{person}{Julian Michael}, \bibinfo{person}{Felix Hill},
  \bibinfo{person}{Omer Levy}, {and} \bibinfo{person}{Samuel Bowman}.}
  \bibinfo{year}{2018}\natexlab{}.
\newblock \showarticletitle{{GLUE: A Multi-Task Benchmark and Analysis Platform
  for Natural Language Understanding}}. In
  \bibinfo{booktitle}{\emph{Proceedings of the 2018 EMNLP Workshop BlackboxNLP:
  Analyzing and Interpreting Neural Networks for NLP}}.
  \bibinfo{publisher}{Association for Computational Linguistics},
  \bibinfo{address}{Stroudsburg, PA, USA}, \bibinfo{pages}{353--355}.
\newblock
\urldef\tempurl%
\url{https://doi.org/10.18653/v1/W18-5446}
\showDOI{\tempurl}


\end{thebibliography}

\end{document}